\documentclass[10pt,twocolumn,letterpaper]{article}
\pdfoutput=1
\usepackage{cvpr}
\usepackage{soul}
\usepackage{xcolor}
\usepackage{wrapfig}
\usepackage{mathtools}
\usepackage{epsfig}
\usepackage{amsmath}
\usepackage{amssymb}
\usepackage{mathtools}
\usepackage{microtype}
\usepackage{wrapfig}
\usepackage{xcolor}
\usepackage{layouts}
\usepackage{mathtools}
\usepackage{printlen}
\usepackage{balance}
\usepackage{graphicx,import}
\usepackage{sepfootnotes}
\usepackage{colortbl}
\usepackage{balance}

\newcommand{\circled}[1]{\raisebox{.5pt}{\textcircled{\raisebox{-.9pt}{#1}}}}
\usepackage{tikz}
\newcommand*\numcircle[1]{\tikz[baseline=(char.base)]{
            \node[shape=circle,draw,inner sep=2pt] (char) {#1};}}
\definecolor{myorange}{HTML}{ED9F7E}
\definecolor{mygreen}{HTML}{61B747}
\newcommand{\best}{\cellcolor{orange!15}}
\graphicspath{{./assets/}}

\definecolor{cvprblue}{rgb}{0.21,0.49,0.74}
\usepackage[pagebackref,breaklinks,colorlinks,allcolors=cvprblue]{hyperref}
\def\paperID{****}
\def\confName{CVPR}
\def\confYear{2025}

\newcommand{\energy}{\mathcal{E}}

\newcommand{\de}{\partial}

\newcommand{\image}{\mathcal{I}}

\newcommand{\hist}{\mathcal{P}}

\newcommand{\histb}{\mathcal{H}}

\newcommand{\R}{\mathbb{R}}

\newcommand{\x}{\mathbf{x}}

\newcommand{\y}{\mathbf{y}}

\definecolor{r}{HTML}{EF625C}
\definecolor{y}{HTML}{E3B67D}

\newtheorem{result}{Result}

\title{Locally Orderless Images for Optimization in Differentiable Rendering}

\author{Ishit Mehta \qquad Manmohan Chandraker \qquad Ravi Ramamoorthi \\
\\
University of California San Diego}

\begin{document}
\maketitle
\begin{abstract} 
    Problems in differentiable rendering often involve optimizing scene
    parameters that cause motion in image space. The gradients for such
    parameters tend to be sparse, leading to poor convergence. While existing
    methods address this sparsity through proxy gradients such as topological
    derivatives or lagrangian derivatives, they make simplifying assumptions
    about rendering. Multi-resolution image pyramids offer an alternative
    approach but prove unreliable in practice. We introduce a method that uses
    locally orderless images --- where each pixel maps to a histogram of
    intensities that preserves local variations in appearance. Using an inverse
    rendering objective that minimizes histogram distance, our method extends
    support for sparsely defined image gradients and recovers optimal
    parameters. We validate our method on various inverse problems using both synthetic and real data.
\end{abstract}

\section{Introduction}
\label{sec:intro}

Much of the recent work addresses inverse rendering with analysis-by-synthesis
--- start with an initial guess of scene parameters, render an image using a
differentiable renderer~\cite{laine20,li18,loper2014opendr,mitsuba3,ravi20},
compare it with a given photograph, estimate gradients, and update the
parameters iteratively.  While the success of gradient-based optimization in
machine learning validates this approach, significant challenges remain. Despite
having algorithms that accurately estimate gradients for the physics of image
formation, optimization hurdles such as local minima, noisy loss landscapes, and
the search for good initialization and parameterization all still
persist~\cite{metz2021gradients}.  

\begin{figure}
    \small 
    \def\svgwidth{\columnwidth}
    \hypersetup{citecolor=white}
    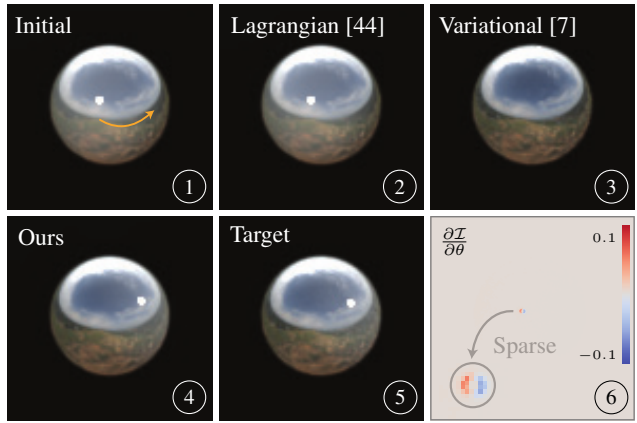
    \hypersetup{citecolor=cvprblue}
    \caption{
    \label{fig:teaser}
        Image gradients are sparse with respect to optimization parameters that
        induce motion in the image space \circled{6}. We show an inverse problem
        with the goal of recovering the position ($\theta$) of a distant light
        source from a synthetic image of a shiny ball, \ie \circled{1} to
        \circled{5}. Existing methods compute proxy gradients such as:
        Lagrangian derivatives~\cite{xingDifferentiableRenderingUsing2022},
        which track only primary-ray intersections, or variational
        derivatives~\cite{fischer2023plateau}, which can be prone to local
        minima. Our method uses standard RGB gradients and uses an inverse
        rendering objective that matches locally orderless images.
        }
\end{figure}

For many inverse problems, the ideal optimization trajectory requires long-range
motion of pixels and image features. Consider the example shown in
Fig.~\ref{fig:teaser}, where a shiny ball is lit by an unknown light source that
we wish to recover from a given image. A primary effect of optimizing the light
position here is on the motion of the specular highlight on the ball. Since
image gradients are sparse (Fig.~\ref{fig:teaser} \circled{6}) for such
parameters, the optimization landscape is rife with local
minima and plateau regions~\cite{mehta2023topological,fischer2023plateau}.
Existing methods address the gradient-sparsity problem with proxy
gradients such as topological derivatives~\cite{mehta2023topological},
lagrangian derivatives~\cite{xingDifferentiableRenderingUsing2022}, and
variational derivatives~\cite{fischer2023plateau} --- all of which are either
expensive to
compute~\cite{fischer2023plateau,xingDifferentiableRenderingUsing2022}, only
work with implicit geometry~\cite{mehta2023topological}, or are restricted to
primary light-transport
effects~\cite{mehta2023topological,xingDifferentiableRenderingUsing2022}.  
We propose a complementary approach that is compatible with RGB gradients from a
differentiable renderer~\cite{li2020diffvg,mitsuba3,laine20} --- which we
demonstrate to be expressive for a variety of inverse problems that have
multiple locally-optimal solutions (as in Fig.~\ref{fig:teaser}).  

Our method builds on scale-space matching~\cite{witkin1987scale}, where signals
are matched at multiple resolutions to measure similarity.  In differentiable
rendering, multi-scale matching has received limited attention, with a few
works~\cite{fischer2023plateau,xingDifferentiableRenderingUsing2022} noting its
unreliability. In this work, we observe that this unreliability stems from the
standard approach of using multi-resolution images or Gaussian
Pyramids~\cite{adelson1984pyramid}, which average out local appearance and
geometric details. Instead, we repurpose the inverse rendering objective to
match local histograms that preserve the \emph{full} distribution of intensity
values within a neighborhood. 
Originally proposed by Griffin~\cite{griffin1997scale} to address imprecision in
radiance measurements, and later extended by Koenderink and van
Doorn~\cite{koenderink1999structure} for reasoning about image topology --- we
postulate that the idea of viewing images as a family of histograms might be
relevant in the seemingly unrelated field of inverse rendering.
\begin{figure}
    \def\svgwidth{\columnwidth}
    \small
\begingroup%
  \makeatletter%
  \providecommand\color[2][]{%
    \errmessage{(Inkscape) Color is used for the text in Inkscape, but the package 'color.sty' is not loaded}%
    \renewcommand\color[2][]{}%
  }%
  \providecommand\transparent[1]{%
    \errmessage{(Inkscape) Transparency is used (non-zero) for the text in Inkscape, but the package 'transparent.sty' is not loaded}%
    \renewcommand\transparent[1]{}%
  }%
  \providecommand\rotatebox[2]{#2}%
  \newcommand*\fsize{\dimexpr\f@size pt\relax}%
  \newcommand*\lineheight[1]{\fontsize{\fsize}{#1\fsize}\selectfont}%
  \ifx\svgwidth\undefined%
    \setlength{\unitlength}{234.38543845bp}%
    \ifx\svgscale\undefined%
      \relax%
    \else%
      \setlength{\unitlength}{\unitlength * \real{\svgscale}}%
    \fi%
  \else%
    \setlength{\unitlength}{\svgwidth}%
  \fi%
  \global\let\svgwidth\undefined%
  \global\let\svgscale\undefined%
  \makeatother%
  \begin{picture}(1,0.52624103)%
    \lineheight{1}%
    \setlength\tabcolsep{0pt}%
    \put(0,0){\includegraphics[width=\unitlength,page=1]{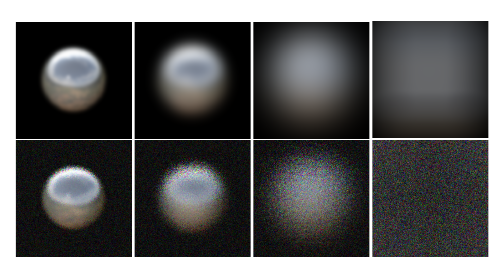}}%
    \put(0.01621589,0.29400187){\color[rgb]{0,0,0}\rotatebox{90}{\makebox(0,0)[lt]{\lineheight{1.25}\smash{\begin{tabular}[t]{l}$\sigma$ space\end{tabular}}}}}%
    \put(0.34199787,0.51642486){\color[rgb]{0,0,0}\makebox(0,0)[lt]{\lineheight{1.25}\smash{\begin{tabular}[t]{l}\footnotesize{Increasing Kernel Width}\end{tabular}}}}%
    \put(0.01838857,0.05495432){\color[rgb]{0,0,0}\rotatebox{90}{\makebox(0,0)[lt]{\lineheight{1.25}\smash{\begin{tabular}[t]{l}$\alpha$ space\end{tabular}}}}}%
    \put(0,0){\includegraphics[width=\unitlength,page=2]{fig5.pdf}}%
  \end{picture}%
\endgroup%

\caption{
\label{fig:spaces} \textbf{Inner and Extent scale spaces.} The image
representation $\hist(\x,k,\alpha,\beta,\sigma)$, is composed of three distinct
histogram-valued scale spaces. The $\sigma$-space (\emph{top}) controls the
effective resolution of the image and the $\alpha$-space (\emph{bottom}) defines
the spatial extent of histogram integration. The rendered images shown here are
intensities sampled as $\image(\x)=k, k\sim\hist(\x, k, \theta,
\alpha,\beta,\sigma)$ for given kernel parameters $\alpha$ and $\sigma$, and bin
width $\beta$. We recover inverse rendering parameters $\theta$ by matching
these locally orderless structures for
rendered and reference images.} 
\end{figure}

Formally, we render images in three distinct scale
spaces~\cite{koenderink1999structure}: 1) the inner ($\sigma$) scale, describing
the effective resolution of the image (Fig.~\ref{fig:spaces} \emph{top}), 2) the
tonal ($\beta$) scale, which quantifies the imprecision in measured radiance at
a location ($\x$), and 3) the extent ($\alpha$) scale, which relates to the
spatial region over which a histogram is computed (Fig.~\ref{fig:spaces}
\emph{bottom}). The resulting representation, $\hist(\x, \alpha, \beta,
\sigma)$, is considered \emph{locally orderless}~\cite{koenderink1999structure}
as the histograms eliminate any spatial variations in intensity, preserving only
their distribution. We find that matching images within this locally orderless 
structure extends gradient support (Fig.~\ref{fig:sigma_scale}) and is effective
in solving inverse rendering problems that require long-range motion of image
features (Fig.~\ref{fig:main_results}). Our method is straightforward to
implement, requires no modifications to a differentiable renderer, is robust
under noise, and works with arbitrary geometry representations and complex
light-transport effects.

Overall, our method achieves good local minima for diverse optimization problems
arising in differentiable rendering (\S~\ref{sec:exp}). Results on inverse
vectorization (\S~\ref{exp:vector}), path tracing (\S~\ref{exp:path}) and
rasterization (\S~\ref{exp:raster}) validate that our method can reliably use
standard RGB gradients, while methods that require computing additional
gradients may fail. Lastly, we show that our method is also compatible with
variational optimization on scenes with complex-light transport effects
(\S~\ref{exp:variational}) and real data (Fig.~\ref{fig:real}).

\section{Related Work}
\label{sec:related} 
The goal of differentiable rendering is to compute derivatives of the rendering
integral~\cite{kajiyaRENDERINGEQUATION1986} with respect to scene parameters.
Estimating the rendering derivative has a long history in vision and
graphics~\cite{arvo1994irradiance,ramamoorthi2007first,ward1992irradiance}, and
proves useful for accurate shape estimation~\cite{nayar1991},
inverse scattering~\cite{gkioulekas2013inverse}, material
acquisition~\cite{khungurn2015matching,bousseau2011optimizing}, and various
vision tasks~\cite{barron2015shape}. While early works
focus on deriving gradients for specific applications, more recent works by Loper
and Black~\cite{loper2014opendr}, Li~\etal~\cite{li18},
Jakob~\etal~\cite{mitsuba3}, Laine~\etal~\cite{laine20}, and
Ravi~\etal~\cite{ravi20} propose general-purpose differentiable renderers.
These renderers come with different tradeoffs, however we focus specifically on
maximizing the utility of the estimated gradients for solving inverse problems.

In contrast to neural networks, where parameter gradients can have global
support~\cite{ia2016deep}, rendering gradients are more local, especially for
parameters that define visibility in the scene. Optimization using gradient
descent is difficult in this case, as shown
in~\cite{mehta2023topological,xingDifferentiableRenderingUsing2022,fischer2023plateau}.
Existing methods extend gradient support either by using alternate geometry
gradients~\cite{mehta2023topological,xingDifferentiableRenderingUsing2022} or
through the notion of \emph{differentiating through blurring} --- either by
blurring discontinuities~\cite{liu2019soft,yariv2021volume,wang2024simple,fischer2024zerograds} or
blurring the parameter space~\cite{fischer2023plateau} as a form of stochastic
approximation~\cite{deliot2024transforming} of the gradient. Our approach
operates entirely in image space and uses scale-space
matching~\cite{witkin1987scale}. While this makes our method conceptually
orthogonal to other methods, we show that they are complementary
(\S~\ref{exp:variational}).

Scale-space matching remains relatively underexplored in differentiable
rendering. While Li~\etal~\cite{li18} use Gaussian Pyramids to avoid local
minima when optimizing geometry primitives, and
Vicini~\etal~\cite{vicini2022differentiable} apply them for recovering signed
distance functions, Xing~\etal~\cite{xingDifferentiableRenderingUsing2022} show
that using multi-resolution images is unreliable. In \S~\ref{sec:method}, we
illustrate why using image pyramids may fail. Our approach uses three distinct
scale spaces implemented as Locally Orderless Images (LOIs), as proposed by
Koenderink and van Doorn~\cite{koenderink1999structure}. Unlike Gaussian
pyramids that only match mean intensities within neighborhoods, our method
matches entire distributions, leading to better recovery. LOIs have proven
effective for various vision tasks~\cite{van2000applications}, including image
retrieval~\cite{lazebnik2009spatial}, object tracking~\cite{oron2015locally},
and non-linear filtering~\cite{van2001local,van2001color}. 

\section{Method}
\label{sec:method} 

Consider an image $\image(\x; \theta):\R^2\mapsto\R$ as a set of direct or
indirect observations of parameters $\theta$ that model the geometry and the
appearance of a scene. Starting from an unknown set of parameters $\theta$, the
goal is to recover an optimal set such that the rendered image matches
identically to a given reference. We use a differentiable renderer to compute
derivatives $\frac{\de\image}{\de\theta}$ and minimize an error function
$\energy$ through gradient descent. As shown in the example in
Fig.~\ref{fig:teaser}, image gradients can be sparse and ineffective for inverse
problems~\cite{mehta2023topological,xingDifferentiableRenderingUsing2022,fischer2023plateau,li18}.

As an illustration, consider the 1D toy example shown in
Fig.~\ref{fig:sigma_scale} A reference image shows a disk for which we
wish to recover its position $\theta$ along the horizontal axis. For any
initialization, the image gradients are sparsely localized at the disk's
silhouette. In the very likely case of no overlap between the initial and the
target disks, $\frac{\de\energy}{\de\theta}=0$, providing no signal to update
the initial choice of $\theta$. Even with some overlap between the two
disks, $\frac{\de\energy}{\de\theta}$ is non-zero but still sparse, making the
optimization sensitive to noise.

\paragraph{Inner Scale}
To increase gradient support, we render images at different
resolutions~\cite{witkin1987scale} and match them in scale
space~\cite{witkin1987signal}. We apply a linear filter with a progressively
increasing kernel width $\sigma\in\R_+$ to images rendered at their stationary
resolution ($\sigma = 0$):
\begin{figure}
    \def\svgwidth{\columnwidth}
    \footnotesize
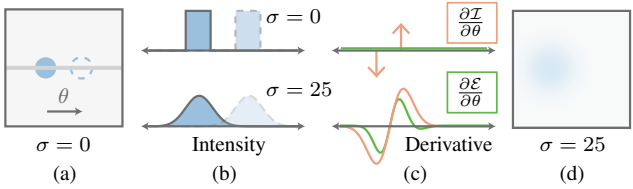
\caption{
\label{fig:sigma_scale}
    \textbf{Scale-space matching extends gradient support.} Given an image (a)
    of a disk we recover its position $\theta$ on the horizontal axis. At
    stationary resolution ($\sigma=0$), the initial and target (dotted) disks do
    not overlap, as shown in the corresponding 1D signals in (b). The image
    gradient $\frac{\de\image}{\de\theta}$ is sparse
    (\textcolor{myorange}{orange}) and is non-zero only at the boundaries of the
    disk (c-\emph{top}). The error gradient $\frac{\de\energy}{\de\theta}$ is zero
    everywhere (\textcolor{mygreen}{green}) and the optimization is stuck in a
    local minimum. When matching at coarser scales (d), the gradients
    are no longer sparse (c-\emph{bottom}), leading to optimal recovery.}
\end{figure}

\begin{equation}
\image(\x; \theta, \sigma) = (G \ast I)(\x; \sigma),
\end{equation}
\\
where $\ast$ is the convolution operator and $G$ is the Gaussian aperture function
that defines the \emph{inner scale}:

\begin{equation}
    \label{eq:Gaussian}
G(\x; \sigma) = \frac{1}{\sqrt{2\pi\sigma^2}}\exp\left(-\frac{\x\cdot\x}{2\sigma^2}\right), \quad \sigma > 0.
\end{equation}

Let us revisit the example in Fig.~\ref{fig:sigma_scale}. At coarser scales
($\sigma > 0$), we find that the image gradients have an extended support with a
strong enough signal for convergence as the disks turn into larger overlapping
blobs. This example reveals a key insight: while gradients are inherently local,
blurring images with a scale-space kernel increases their support and simplifies
the problem of alignment in image space. Note that this idea of \emph{gradient
diffusion} is similar to variational
optimization~\cite{fischer2023plateau,staines2012variational,suh2022differentiable},
where a blur kernel is alternatively used in the \emph{parameter space} to
achieve a similar effect of extending gradient support and enabling long-range
matching in image space.

\begin{figure}
    \def\svgwidth{\columnwidth}
    \footnotesize
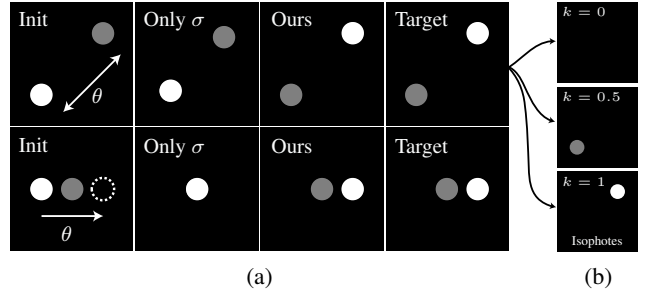
\caption{
\label{fig:tonalSeparation}
    \textbf{Tonal Separation.} Shown are two (a-\emph{top} and a-\emph{bottom})
    1D inverse problems where we recover disk positions ($\theta$) from images
    (\emph{left}). Image matching within $\sigma$-space measures only the errors
    in the mean of the intensity distributions at each scale. In inverse
    settings that involve multiple objects with different appearances, this
    approach is likely to get stuck in a local minimum (a-\emph{center-left}).
    The $\alpha$-space integration kernels are intensity-aware and treat images
    as sets of distinct equal-intensity isophotes (b). When images are matched
    in all three scale spaces, the optimization is less prone to getting stuck in local minima
    (a-\emph{center-right}).}
\end{figure}

\begin{figure}
    \def\svgwidth{\columnwidth}
    \footnotesize
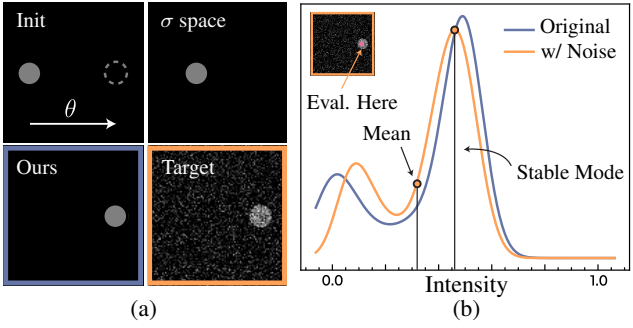
\caption{
\label{fig:noise} 
\textbf{Histogram matching is less sensitive to noise.}
To recover the position ($\theta$) of a circular disk from a noisy reference
image (a-\emph{bottom-right}), methods that match images only at their
stationary resolution or in $\sigma$-scale space fail --- as they overlook
imprecision and uncertainty in radiance measurements. Our method uses a tonal
parameter ($\beta$) to account for intensity uncertainty and an extent
scale-space to preserve the distribution modes at coarser scales (b), leading to
optimal recovery of $\theta$.}
\end{figure}

Scale-space matching performs well for simple, controlled problems, but it
remains unreliable in more realistic scenarios.  As shown in recent
work~\cite{xingDifferentiableRenderingUsing2022,fischer2023plateau}, for inverse
problems with complex geometry and light-transport effects, blurring and
multi-scale strategies are not robust.  We identify two primary reasons for
this.  \emph{First}, blurring has an averaging effect that changes the local
appearance
--- this makes matching image features more difficult. \emph{Second}, blurring suppresses
high-frequency details. With minor tweaks to the previous toy example, such as
including image noise or multiple objects (Fig.~\ref{fig:tonalSeparation} and
\ref{fig:noise}), we find that inner-scale matching still suffers from
local minima problems. Our method uses two additional scale parameters to
achieve optimal recovery.

\paragraph{Tonal Scale}
We use scale-imprecision space~\cite{griffin1997scale} and relax the
assumption of using images as real-valued functions. Instead, at each pixel
location, we model a radiance distribution that captures the uncertainty in the
measurements. Even in ideal settings, this is reasonable, as pixels integrate
light over a non-zero area and measurements have finite
precision~\cite{poincare1905science}. In the context of inverse rendering,
images can be noisy due to Monte Carlo
integration~\cite{kajiyaRENDERINGEQUATION1986} or external factors such as
sensors and lenses. Explicitly modeling uncertainty in estimated parameters can
lead to better recovery, as also observed in~\cite{zhou2024estimating}. Our
method models the uncertainty in radiance estimates using 1D kernels with
bandwidth $\beta\in\R_+$. The result is a family of spatially-varying
probability distributions $\hist$ defined on the image plane:

\vspace{-1em}
\begin{equation}
    \label{eq:beta}
\hist(\x, k; \theta, \sigma, \beta) = \frac{1}{\sqrt{2\pi\beta^2}} 
\exp\left(-\frac{\scriptstyle(k - \image(\x; \theta, \sigma))^2}{\scriptstyle2\beta^2}\right), 
\end{equation}
where $\beta$ is the scale in intensity domain and controls the tonal
resolution, and $k\in\left[0, 1 \right]$ is intensity. For a given $k$,
$\hist(\x, k)$ is a soft \emph{isophote} that measures the probability of
intensity being $k$ at $\x$. This approach provides the benefit of \emph{tonal
separation}, where each isophote is treated as a distinct image, leading to
improved separation both in visual features and image gradients. The inverse
problem in Fig.~\ref{fig:tonalSeparation} illustrates the benefit of using a
tonal scale.
In practice, we use $\beta$
as the width of the bins used for discretizing $\hist$ as histograms.

\paragraph{Extent Scale}
To capture spatial relationships between the local distributions $\hist$, we can
extend this representation further using the idea of locally orderless
images~\cite{koenderink1999structure}. Similar to integrating image intensity
over a spatial extent as in $\sigma$ scale space, we use an aperture function
$A$ that integrates the locally-defined histograms as follows:
\vspace{-0.5em}
\begin{equation}
    \label{eq:loi}
\histb(\x, k; \theta, \sigma, \beta, \alpha) = \int_{\R^2} A(\x-\y; \alpha) \hist(\y, k;
\theta, \sigma, \beta)\ d\y,
\end{equation}
\\
where $A$ is similar to the kernel defined in Eq.~\ref{eq:Gaussian}. Unlike
conventional scale-space blurring which operates directly on radiance values,
Eq.~\ref{eq:loi} blurs histogram contributions. This distinction enables the
method to preserve the effective \emph{modes} of the radiance distribution
across scales --- thereby retaining both the appearance and geometric
characteristics of the image even at coarser resolutions. Consequently, our
method is well-suited to handle perturbations in the reference images as the
modes of the intensity distribution remain stable under
noise~\cite{van2001local}. See Fig.~\ref{fig:noise} for an illustrative example.

\paragraph{Histogram Matching}
Equipped with three separate scale spaces, defined using the parameters $\alpha,
\beta$ and $\sigma$, we can now pose the inverse rendering objective as
minimizing the difference between the distributions of rendered ($\histb'$) and
reference ($\histb^{\text{gt}}$) images. Matching histograms and computing
related distance metrics is a well-studied task in graphics and optimal
transport literature~\cite{pele2010quadratic,peyre2019computational}.  We use
the Wasserstein distance which has a closed form for 1D distributions. At given
scales $\alpha, \beta$ and $\sigma$, the distance between two histograms is
estimated as the summation of point-wise errors between their cumulative
distribution functions (cdf):
\small
\begin{equation}
    \label{eq:wass}
    \begin{aligned}
    \energy(\theta, \alpha, \beta, \sigma) = \int_{\R^2}\int_0^1 & \left[\text{cdf}_{\histb'}(\x, k; \theta, \alpha, \beta, \sigma) \right. \\
    & \left. -\text{cdf}_{\histb^{\text{gt}}}(\x, k; \alpha, \beta, \sigma)\right]^{1/p}\ dkd\x,\\
    \text{where,}\quad \text{cdf}_{\histb}(\x,k;\cdot) &= \int_0^k\histb(\x, j;\cdot)\ dj
    \end{aligned}
\end{equation}
\normalsize
We compute the total error as the summation over all the scales, \ie
$\energy_{\text{total}}(\theta) =
\sum_{\alpha,\beta,\sigma}\energy(\theta,\alpha,\beta,\sigma)$. We use $p=1$ in our experiments.

\begin{figure}
    \def\svgwidth{\columnwidth}
    \footnotesize
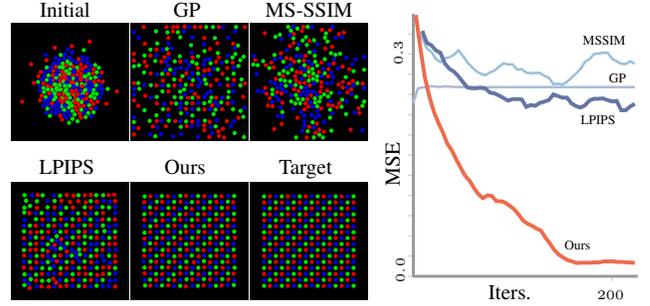
\caption{
    \label{fig:disks}
    \textbf{Comparisons with multi-scale methods.} Given an image we optimize
    positions of $256$ disks starting from random intialization. We use
    diffvg~\cite{li2020diffvg} to compute RGB gradients. Our method recovers the
    closest optimal arrangement. The error plot on the right shows mean-squared
    error (MSE) between rendered and target images \vs iterations.
    }
\end{figure}

\section{Results}
\label{sec:exp}
We evaluate our method using three differentiable renderers:
diffvg~\cite{li2020diffvg} for vector graphics, a path tracer~\cite{mitsuba3},
and a rasterizer~\cite{laine20}. Our experiments are designed for three goals:
First, we compare our method against other multi-scale approaches like Gaussian
Pyramids~\cite{adelson1984pyramid} and MS-SSIM~\cite{wang2003multiscale}
(\S~\ref{exp:vector}). Second, we evaluate how parameter-space blurring
(PRDPT~\cite{fischer2023plateau}) performs compared to scale-space matching
(\S~\ref{exp:path}). Third, we test the reliability of RGB gradients versus
proxy gradients like RGBXY~\cite{xingDifferentiableRenderingUsing2022}.  We also
show how our method complements parameter-space blurring through two example
scenes (\S~\ref{exp:variational}), including one with real data. 

\subsection{Differentiable Vectorization}
\label{exp:vector}
\begin{table}[t]
    \centering
    \caption{
        \label{tab:disks}
        \textbf{Quantitative comparisons for 2D.} We recover positions of $n=[4,
        16, 32, 64, 256]$ disks with \emph{known} color from a reference
        $128\times 128$ image. Image gradients are computed using
        diffvg~\cite{li2020diffvg} and sparsely defined along the silhouettes of
        the disks~\cite{mehta2023topological}. Compared to other scale-space
        approaches, we find our method to be best suited for this task. 
    }
    \resizebox{\columnwidth}{!}{
            \begin{tabular}{@{}r@{\hskip 1em}cc|cc|cc|cc|cc@{}}
                \toprule
                $n$ Disks & \multicolumn{2}{c|}{4} & \multicolumn{2}{c|}{16} & \multicolumn{2}{c|}{32} & \multicolumn{2}{c|}{64} & \multicolumn{2}{c}{256} \\
                \cmidrule(l){2-11}
                Method & PSNR & SSIM & PSNR & SSIM & PSNR & SSIM & PSNR & SSIM & PSNR & SSIM \\
                \midrule
                GP~\cite{adelson1984pyramid} & 25.05 & 0.97 & 19.83 & 0.91 & 16.83 & 0.83 & 15.34 & 0.76 & 9.60 & 0.34 \\
                MS-SSIM~\cite{wang2003multiscale} & 27.44 & 0.98 & 21.41 & 0.94 & 17.93 & 0.87 & 15.42 & 0.78 & 9.83 & 0.35 \\
                LPIPS~\cite{zhang2018perceptual} & 27.79 & 0.98 & 20.94 & 0.93 & 17.88 & 0.86 & 17.04 & 0.82 & 15.2 & 0.76 \\
                Ours & \best 30.40 & \best 0.99 & \best 33.55 & \best 0.99 & \best 27.43 & \best 0.97 & \best 28.23 & \best 0.96 & \best 21.57 & \best 0.90 \\
                \bottomrule
                \end{tabular}
    }
    \vspace{-1em}
\end{table}

Multi-scale feature matching is widely used in vision and graphics for various
tasks. Three prominent techniques include: (a) the Gaussian
pyramid~\cite{adelson1984pyramid}, which approximates the $\sigma$-scale space
and matches the mean of intensity distributions across scales; (b) multi-scale
structural similarity (MS-SSIM)~\cite{wang2003multiscale}, which matches both
the mean and covariance of the distributions; and (c)
LPIPS~\cite{zhang2018perceptual}, which uses a hierarchy of features from a deep
neural network to measure perceptual similarity. Our approach, using three
distinct scale spaces, is similar to these methods but is specifically designed
for inverse rendering problems.

We design a test benchmark to evaluate multi-scale methods. Similar to previous
experiments, the objective is to recover the positions of 2D disks with known
colors from a given image. This is a representative task for inverse rendering
problems that require long-range feature matching with sparse gradient
information. The benchmark includes five problem sets, with difficulty
determined by the number of disks in each image. Note that this a much more
difficult task than optimizing color, since appearance gradients are not as
sparse as geometry gradients~\cite{mehta2023topological}. Optimization
parameters are randomly initialized, while target positions form a uniform grid.
Differentiable vectorization~\cite{li2020diffvg} is used to compute image
gradients. For other multi-scale approaches, we measure the L2 error between
features at different scales, while our method uses Wasserstein distance
(Eq.~\ref{eq:wass}) to compare histograms. The reference images are $128\times
128$ at stationary resolution and we choose $\alpha=[1, 5, 15], \sigma=[1, 5,
15, 45]$ and $\beta=0.125$ as the scale space parameters. 
For consistency, we
perform $10$ runs, each with a different initialization, and report mean PSNR
and SSIM in Table~\ref{tab:disks}. The optimized images for $n=256$ are shown in
Fig.~\ref{fig:disks} along with the corresponding error plot. 

\subsection{Differentiable Path Tracing}
\label{exp:path}
\begin{figure*}
    \def\svgwidth{0.95\textwidth}
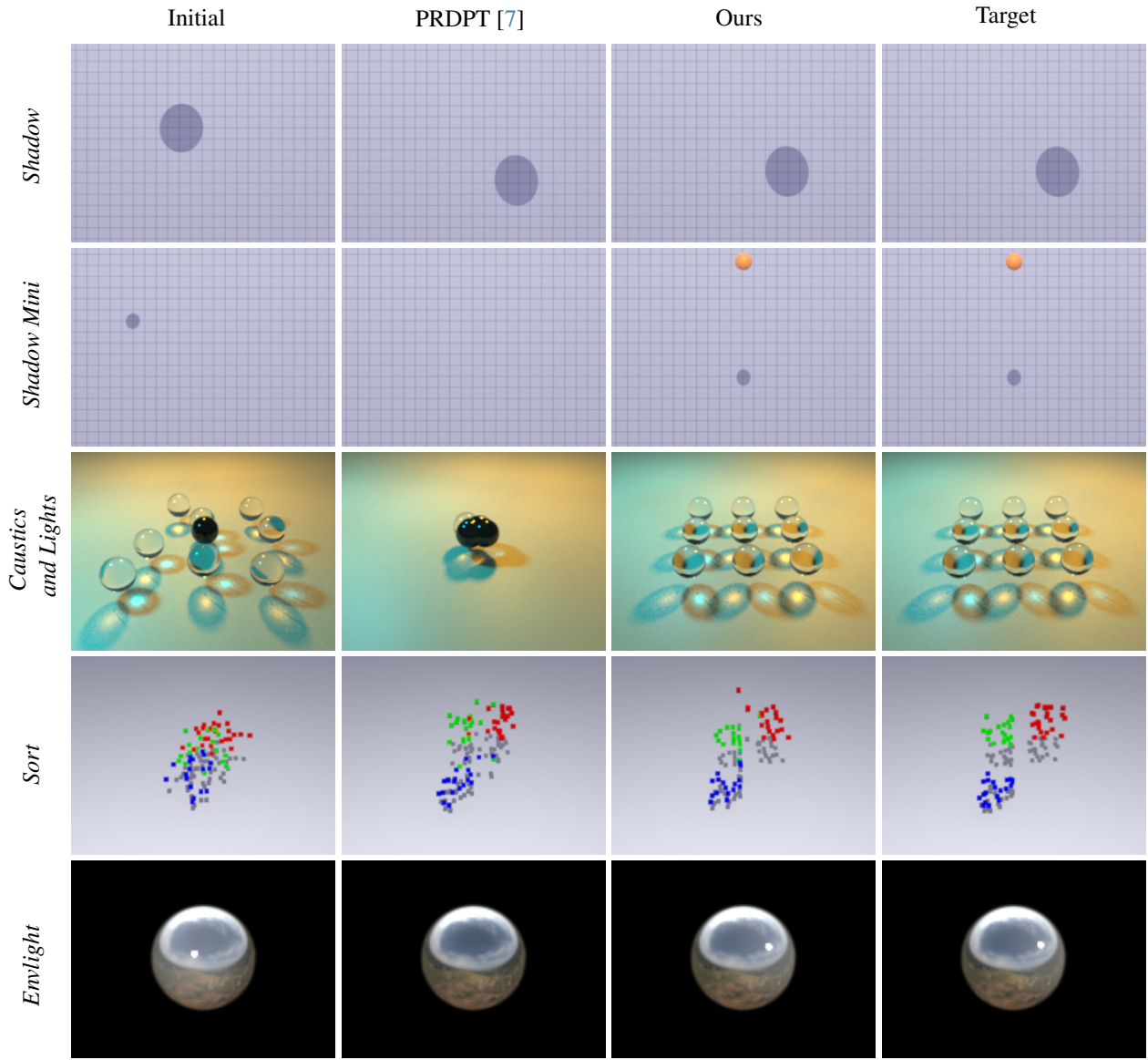
\caption{
    \label{fig:main_results}
    \textbf{Quantitative comparisons with parameter-space blurring.} We show
    five separate scenes (one in each row) with varying degrees of difficulty.
    Starting from a random initialization (left-most column), the goal is to
    recover the positions of different primitives so that the rendered image
    matches the reference (right-most column). Compared to parameter-space
    blurring (PRDPT~\cite{fischer2023plateau}), we find our method that uses
    image-space local histogram matching is more suitable for these tasks. For
    quantitative comparisons, refer to Table~\ref{tab:main_results}.
    }

\end{figure*}

\begin{table*}
    \centering
    \caption{
        \label{tab:main_results}
        We evaluate our method and PRDPT~\cite{fischer2023plateau}, which
        employs parameter-space blurring, across five synthetic scenes of
        varying difficulty. We also present results using other multi-scale
        baselines. Our method recovers the optimal parameter configuration for
        all scenes, despite relying on sparsely defined RGB gradients.
        }
    \resizebox{\textwidth}{!}{ \begin{tabular}{rcccccccccc}
        \toprule
        & \multicolumn{2}{c}{\emph{Shadow}} & \multicolumn{2}{c}{\emph{Shadow
        Mini}} & \multicolumn{2}{c}{\emph{Caustics and Lights}} &
        \multicolumn{2}{c}{\emph{Sort}} & \multicolumn{2}{c}{\emph{Envlight}} \\
        \cmidrule(lr){2-3} \cmidrule(lr){4-5} \cmidrule(lr){6-7} \cmidrule(lr){8-9} \cmidrule(lr){10-11}
        & PSNR & MAE & PSNR & MAE & PSNR & MAE & PSNR & MAE & PSNR & MAE \\
        \midrule
        PRDPT~\cite{fischer2023plateau} & 31.77 & 0.0035 & 27.74 & 6.0147 &
        17.09 & 4.8801 & 22.25 & 0.9201 & 31.87 & 0.7543\\
        Ours & \best 54.33 & \best 0.0034 & \best 54.73 & \best 0.0033 & \best
        32.58 & \best 0.3475 & \best 25.23 & \best 0.4855
        & \best 36.17 & \best 0.0775 \\
        Mitsuba~\cite{mitsuba3} (GP) & 24.44 & 1.1349 & 24.06 & 2.3491 & 21.02 & 0.7317 & 23.55
        & 0.5189 & 28.73 & 0.8072 \\
        Mitsuba~\cite{mitsuba3} (MSSSIM) & 24.35 & 0.9320 & 24.34 & 0.9061 & 10.30 & 86.2286 &
        18.91 & 0.7742 & 30.26 & 0.6703 \\
        \bottomrule
    \end{tabular}
    }
\end{table*}

Following the evaluation framework in~\cite{fischer2023plateau}, we test our
method on five scenes. The goal here is to evaluate methods that use
the notion of scale either in image space or parameter space (\ie variational
optimization).  We use Mitsuba 3~\cite{mitsuba3} as the default renderer with
the \texttt{prb\_reparam} integrator~\cite{vicini2021prb}.  
Quantitative results evaluating visual appearance (PSNR) and parameter recovery (MAE) are in Table~\ref{tab:main_results}. We use the same kernel
parameters, $\alpha=[1, 5, 15, 45], \beta=0.125,$ and $\sigma=[1, 5]$ for all
the scenes.  All images are rendered at $256\times 192$ resolution. We use Adam with a learning rate of $10^{-2}$ for all methods.

\paragraph{Shadow}
We use the \texttt{Shadow} scene from~\cite{fischer2023plateau}. Given an image,
we recover the position of a sphere that is not visible in the view frustum but
projects a shadow onto a visible plane. This is a representative example for
when RGB gradients perform poorly ~\cite{fischer2023plateau}. Using 
gradients from Mitsuba 3~\cite{mitsuba3}, our method recovers the position of
the sphere. Qualitative results for our method and variational optimization are
shown in Fig.~\ref{fig:main_results}. While variational optimization is able to recover the target shadow, other multi-scale methods get stuck in local minima.

\paragraph{Shadow Mini}
\setlength\intextsep{0pt}
\setlength{\columnsep}{5pt}%
\begin{wrapfigure}{r}{0pt}
    \footnotesize
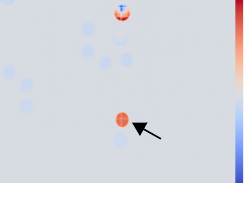
\end{wrapfigure}

We modify the previous scene with two changes: (a) an initialization that is
further away from the target, and (b) changing the scale of the sphere and hence
the area of the shadow. Variational optimization struggles in such scenes.  It
achieves blurring in parameter space through multiple samples that correspond to
images with perturbed positions of the sphere.  Ideally, for this approach to
work, the shadows in the sampled images require some overlap with the target in
the image space to update the parameters reliably. Since that is unlikely to
happen (see inset) with the small shadow footprint in the reference, the
optimization converges to the sphere being pushed out of the scene. The inset
figure plots the difference between the target image and the variational
estimate (avg. image) using $10$ samples. As mentioned, shadows from the
variational samples do not overlap with the target shadow. We test with
different kernel parameters and sampling rates ($\tau$
in~\cite{fischer2023plateau}), all of which perform similarly on this scene.
Since our method operates directly in image space, it reliably recovers the
reference independent of the initial position of the shadows. Other multi-scale
methods converge at different local minima.

\vspace{-1em}
\paragraph{Caustics and Lights}
This scene requires optimization in two separate domains: (a) recovering the
positions of glass spheres, and (b) adjusting lights that are not directly
visible in the view frustum. We normalize the range of both domains within $[-5,
5]$. This task is challenging because, at initialization, the caustic patterns
do not overlap with the reference in either appearance or position. We observe
that matching histograms is particularly suited for this task due to the
presence of Monte Carlo noise in forward rendering and high-frequency caustics
patterns.

\vspace{-1em}
\paragraph{Sort}
In the \texttt{Sort} scene~\cite{fischer2023plateau}, the objective is to
recover the positions of $65$ differently colored primitives from an image. This
involves optimizing the $x$ and $z$ translation coordinates of these primitives,
resulting in a $130$-dimensional optimization problem. Since optimization in
this scene is highly sensitive to the choice of initialization, we perform $20$
different runs with varying random seeds and report the average results in
Table~\ref{tab:main_results}. Comparisons for one of the runs are shown
in Fig.~\ref{fig:main_results}. Since stochastic gradients have high variance
for high dimensional problems~\cite{blei2017variational}, variational
optimization is not ideal for such scenes.

\paragraph{Envlight}
In this scene, we recover the position of a distant light source from an image
of a reflective sphere rendered within a high-frequency environment. The light
source is not directly visible, casting only a small specular highlight on the
sphere, creating an optimization landscape with multiple local minima. Similar
to the \texttt{Shadow Mini} scene, PRDPT~\cite{fischer2023plateau} is stuck in a
local minimum, as one of the sampled images need the highlight to be
exactly aligned with the target. Our method, despite using sparse RGB gradients,
successfully recovers the optimal configuration. Using image pyramids fails at
this task as the high-frequencies and modes of intensity distribution are lost
at coarser scales. 

\subsection{Variational Optimization}
\label{exp:variational} 
In the following two scenes, we demonstrate cases where RGB gradients alone are
not reliable, and show how combining our method with variational
derivatives~\cite{fischer2023plateau} can achieve optimal recovery.

\vspace{-1em}
\paragraph{Kaleidoscope} 
\begin{figure}
    \centering
    \def\svgwidth{0.99\columnwidth}
    \small
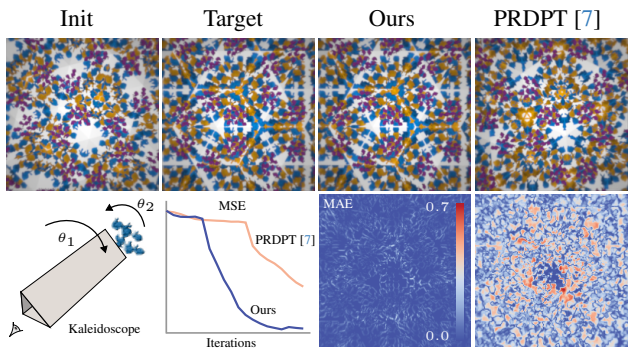
\caption{
    \label{fig:kaleidscope}
    \textbf{Variational Optimization with Locally Orderless Images.} Our method
    works well with other optimization techniques like variational optimization.
    Given an image of a pattern inside a kaleidoscope, we recover the
    configuration of the kaleidoscope ($\theta_1$) and the particles
    ($\theta_2$) reflecting light through it. Using variational derivatives and
    matching images at their stationary resolution leads to sub-optimal
    recovery. By matching local image histograms (Ours), we successfully recover
    the target configuration.
    }
\end{figure}

As shown in~\cite{fipt2023}, RGB gradients are unstable and inefficient for
scenes involving complex light transport. Consider the virtual kaleidoscope
scene shown in Fig.~\ref{fig:kaleidscope}. Our goal is to recover both the
kaleidoscope configuration and the positions of light-reflecting particles from
a single image. The complex light paths within the kaleidoscope result in
high-variance RGB gradients that cannot be reliably used for gradient-based
optimization. Moreover, small parameter perturbations can dramatically change
the kaleidoscope patterns, leading to high variance even in variational
derivatives and preventing optimal recovery. We find that our approach of
combining multi-scale histogram matching with parameter-space blurring is ideal
for this task and successfully recovers the optimal configuration.  We use the
same scale-space parameters as in \S~\ref{exp:path} and render images at
$200\times 200$ resolution.

\begin{table*}\centering 
    \caption{ \label{tab:rasterization} \textbf{Quantitative results with
    differentiable rasterization.} We evaluate our method using gradients
    ($\nabla$) from Nvdiffrast along with other multi-scale methods on test
    scenes provided in~\cite{xingDifferentiableRenderingUsing2022}. We show that
    our method using RGB gradients can reliably recover optimal solutions for
    low-dimensional problems (such as Translation and Rotation) and is ideal for
    higher dimensional problems (such as Camera Pose or Material recovery). Each image is
    rendered at $128\times 128$. We use the scale parameters $\alpha=[1, 5, 15,
    45], \sigma=[0, 5]$ and $\beta = 0.03125$.}
    \resizebox{\textwidth}{!}{ \begin{tabular}{@{}rc@{\hskip
        1em}cc|cc|cc|cc|cc|cc|cc@{}} \toprule & Parameters  &
        \multicolumn{2}{c}{Translation} & \multicolumn{2}{c}{Rotation} &
        \multicolumn{2}{c}{Shape} & \multicolumn{2}{c}{Trans. + Rot.} &
        \multicolumn{2}{c}{Camera Pose} & \multicolumn{2}{c}{Material} &
        \multicolumn{2}{c}{Env. Map} \\ \cmidrule(l){3-16}& Num. views  &
        \multicolumn{2}{c}{6} & \multicolumn{2}{c}{6} & \multicolumn{2}{c}{1}
        & \multicolumn{2}{c}{4} & \multicolumn{2}{c}{1} &
        \multicolumn{2}{c}{6} & \multicolumn{2}{c}{6} \\ \cmidrule(l){3-16} Method &
        $\nabla$ type & PSNR & MAE & PSNR & MAE & PSNR & MAE & PSNR & MAE & PSNR & SSIM & PSNR &
        SSIM & PSNR & SSIM \\ \midrule
        Xing~\etal~\cite{xingDifferentiableRenderingUsing2022} & RGBXY & \best 63.62 & \best
        0.00 & 26.66 & 0.45 & \best 30.75 & \best0.30 & 30.23 & 0.68
        & 23.50 & 0.92 & 33.01 & 0.96 & 32.43 & 0.91 \\
        \midrule
        Ours & RGB & 51.77 & 0.00 & \best 30.73 & \best 0.33 & 28.22 & 0.54 & \best 36.23 & \best
        0.35 & \best 27.22 & \best 0.96 & \best 49.07 & \best 1.00 & \best 55.22
        & \best 1.00 \\
         Nvdiffrast~\cite{laine20} (GP~\cite{adelson1984pyramid}) & RGB & 31.16 & 0.18 & 22.40 & 0.49 & 25.55 & 0.59 &
         19.53 & 2.11 & 14.82 & 0.73 & 42.48 & 0.99 & 47.79 & 0.99 \\
         Nvdiffrast~\cite{laine20} (MS-SSIM~\cite{wang2003multiscale}) & RGB & 23.51 & 0.60 & 17.61 & 0.88 & 11.20 & 0.83 & 17.74
         & 2.37 & 09.01 & 0.33 & 11.19 & 0.72 & 23.30 & 0.73 \\
    \bottomrule \end{tabular} } \end{table*}

\paragraph{Real Scene}
\begin{figure}
    \def\svgwidth{\columnwidth}
    \footnotesize
    \hypersetup{citecolor=white}
    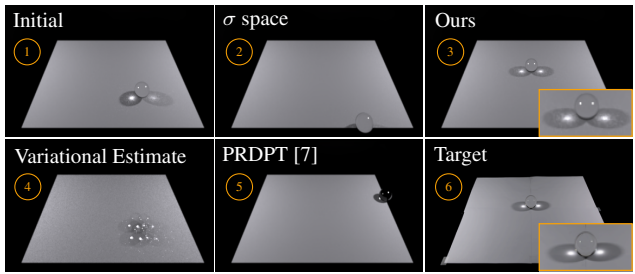
    \hypersetup{citecolor=cvprblue}
    \caption{ 
    \label{fig:real} \textbf{Using a real photograph.} We show a preliminary
    experiment with real data that validates our method's robustness against
    sensor noise and calibration errors. Starting with known material, camera,
    and lighting, and a randomly positioned glass orb \circled{1}, we recover
    the optimal translation from a real photograph \circled{6} using variational
    optimization.  The variational estimate with 8 samples is shown in
    \circled{4}. Matching at stationary resolution \circled{5} or only in the
    $\sigma$ space \circled{2} fails in this task. Our method achieves optimal
    recovery starting from any initial position on the surface.
    \vspace{-1em}
    }
\end{figure}

Inverse rendering with real photographs presents additional challenges, such as
noise from sensors and lenses, and calibration errors. We test our method on a
real scene, where we aim to recover the position of a glass orb on a known
planar surface from a given photograph. The scene is initialized with calibrated
light sources, camera parameters, and known material (borosilicate glass), but
the position of the orb is unknown (Fig.~\ref{fig:real} \circled{1}). Similar to
the previous scene, we use variational optimization.  We observe that using
matching stationary images ($\sigma=0, \alpha=0$) is sensitive to noise and
fails in this task (Fig.~\ref{fig:real} \circled{5}). We attribute this failure
to the same alignment issue as in \texttt{Shadown Mini}, where one of the
samples in the variational estimate (Fig.~\ref{fig:real} \circled{4} shows the
mean of all samples) needs to overlap with the target position of the orb. Our
local-histogram matching approach does not require perfect alignment between the
variational estimate and the reference images, allowing it to reliably recover
the target position of the orb from any initialization on the planar surface
(Fig.~\ref{fig:real} \circled{3}). As an ablation on the $\alpha$ and $\beta$
scale spaces, we also test matching only in the $\sigma$ space with the same
kernel parameters, which fails in this task (Fig.~\ref{fig:real} \circled{2}).

\subsection{Differentiable Rasterization}
\label{exp:raster}
We evaluate our method on scenes provided by
Xing~\etal~\cite{xingDifferentiableRenderingUsing2022}. The tested scenes vary
in complexity, requiring optimization for both geometric and appearance
parameters. Quantitative results are in Table~\ref{tab:rasterization}, using
metrics to assess both visual appearance (PSNR) and parameter recovery accuracy
(MAE). For low-dimensional parameter spaces (e.g.,
\texttt{Translation} and \texttt{Rotation}), our method using \emph{only} RGB
derivatives performs comparably to
Xing~\etal's~\cite{xingDifferentiableRenderingUsing2022} approach, which
utilizes both RGB and Lagrangian derivatives. For high dimensional parameter
spaces, such as optimizing the full camera-pose matrix or texture map, RGB
gradients combined with local histogram matching performs better. We use the
same scale parameters as in \S~\ref{exp:vector}.

\section{Discussion}
\begin{figure}[h!]
    \centering
    \def\svgwidth{0.92\columnwidth}
    \small
\begingroup%
  \makeatletter%
  \providecommand\color[2][]{%
    \errmessage{(Inkscape) Color is used for the text in Inkscape, but the package 'color.sty' is not loaded}%
    \renewcommand\color[2][]{}%
  }%
  \providecommand\transparent[1]{%
    \errmessage{(Inkscape) Transparency is used (non-zero) for the text in Inkscape, but the package 'transparent.sty' is not loaded}%
    \renewcommand\transparent[1]{}%
  }%
  \providecommand\rotatebox[2]{#2}%
  \newcommand*\fsize{\dimexpr\f@size pt\relax}%
  \newcommand*\lineheight[1]{\fontsize{\fsize}{#1\fsize}\selectfont}%
  \ifx\svgwidth\undefined%
    \setlength{\unitlength}{299.28905337bp}%
    \ifx\svgscale\undefined%
      \relax%
    \else%
      \setlength{\unitlength}{\unitlength * \real{\svgscale}}%
    \fi%
  \else%
    \setlength{\unitlength}{\svgwidth}%
  \fi%
  \global\let\svgwidth\undefined%
  \global\let\svgscale\undefined%
  \makeatother%
  \begin{picture}(1,0.27394084)%
    \lineheight{1}%
    \setlength\tabcolsep{0pt}%
    \put(0,0){\includegraphics[width=\unitlength,page=1]{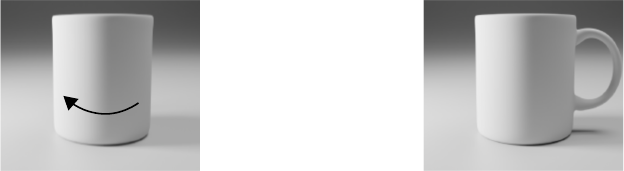}}%
    \put(0.01166762,0.01421146){\color[rgb]{0,0,0}\makebox(0,0)[lt]{\lineheight{1.25}\smash{\begin{tabular}[t]{l}Init\end{tabular}}}}%
    \put(0.69433857,0.01421146){\color[rgb]{0,0,0}\makebox(0,0)[lt]{\lineheight{1.25}\smash{\begin{tabular}[t]{l}Target\end{tabular}}}}%
    \put(0,0){\includegraphics[width=\unitlength,page=2]{inset2.pdf}}%
    \put(0.35203792,0.05931829){\color[rgb]{0,0,0}\makebox(0,0)[lt]{\lineheight{1.25}\smash{\begin{tabular}[t]{l}Mitsuba\\+ Ours\end{tabular}}}}%
    \put(0.15618344,0.05616367){\color[rgb]{0,0,0}\makebox(0,0)[lt]{\lineheight{1.25}\smash{\begin{tabular}[t]{l}$\theta$\end{tabular}}}}%
  \end{picture}%
\endgroup%

\caption{
    \label{fig:mug}
    Our method is not designed to recover parameters that do not influence
the image, such as the rotation of the mug~\cite{fischer2023plateau}.}
\vspace{-1em}
\end{figure}

\label{sec:discussion}
\paragraph{Ablations}
Our method relies on two critical components: the $\alpha$ and $\beta$ scale
spaces. Using only the $\sigma$ space reduces our method to standard Gaussian
Pyramids (GPs), as shown in our quantitative evaluations
(Tables~\ref{tab:disks}, \ref{tab:main_results}, and \ref{tab:rasterization}).
While using only the $\beta$ space preserves tonal variations, it fails to
extend the spatial influence of gradients, leading to poor performance across
our experiments
. 
Since the $\alpha$ space is dependent on $\beta$ (Eq.~\ref{eq:loi}), it cannot
be used in isolation. 
\paragraph{Limitations}
As our method operates exclusively in image space, parameters that do not
influence primary or secondary effects still yield zero gradients in the
LOI representation. For example, in the \texttt{mug} scene (Fig.~\ref{fig:mug})
from~\cite{fischer2023plateau}, optimizing the mug's rotation when its handle
is not visible in the initial view is infeasible. In
such cases, our method is most effective when combined with variational
optimization approaches, where parameter-space blurring can reveal occluded
features like the handle through sampling.

\paragraph{Conclusion}
In this work, we revisit the idea of locally orderless images (LOIs) in the
context of differentiable rendering. One of the key contributions of this work
is bridging these previously disconnected areas in computer vision and graphics.  Through
various experiments, we show that our method yields better recovery in
optimization problems where existing methods are likely to fail. Our method is
straightforward to implement and integrates seamlessly with existing
differentiable renderers without requiring any modifications to their core
functionality.
\paragraph{Acknowledgments}
This work was supported in part by NSF grant 2402583.  We also
acknowledge NSF grant 2110409, a Qualcomm Innovation Fellowship, gifts
from Adobe, Google, Qualcomm and Rembrand, the Ronald L. Graham Chair
and the UC San Diego Center for Visual Computing.

{
    \balance
    \small
    \bibliographystyle{ieeenat_fullname}
    \bibliography{main, zot}
}

\end{document}